\documentclass[twocolumn,aps,reprint,nofootinbib,superscriptaddress]{revtex4-1}

\usepackage{graphicx}
\usepackage{epsfig,amsmath}
\usepackage{amssymb}
\usepackage{rotate}
\usepackage{color}
\usepackage{hyperref}
\usepackage{bm}

\newcommand{\be}{\begin{eqnarray}}
\newcommand{\ee}{\end{eqnarray}}
\newcommand{\Pnl}{P^\mathrm{nl}}
\newcommand{\Plin}{P^\mathrm{lin}}
\newcommand{\Pspt}{P^\mathrm{SPT}_{1-loop}}
\newcommand{\PH}{P^\mathrm{Hf}}

\newcommand{\Omegam}{{\Omega_\mathrm{m}}}

\newcommand{\e}{{\mathrm{e}}}

\newcommand{\dd}{{\rm d}}

\newcommand{\vk}{{\bf k}}
\newcommand{\vq}{{\bf q}}

\newcommand{\Dirac}{\delta_{\rm D}}

\newcommand{\neff}{n_{\mathrm{eff}}}
\newcommand{\rnl}{R_{\mathrm{nl}}}

\newcommand{\rsmooth}{R^{\mathrm{smooth}}}
\newcommand{\rdelta}{R^{\delta}}

\begin{document}
\title{Failures of Halofit model for computation of Fisher Matrices}

\author{P. Reimberg}
\affiliation{Sorbonne Universités, CNRS, UMR 7095, Institut d'Astrophysique de Paris, 98 bis bd Arago, 75014 Paris, France}
\email{reimberg@iap.fr}
\author{F. Bernardeau}
\affiliation{Sorbonne Universités, CNRS, UMR 7095, Institut d'Astrophysique de Paris, 98 bis bd Arago, 75014 Paris, France}
\affiliation{CEA - CNRS, URA 2306, Institut de Physique Théorique, F-91191 Gif-sur-Yvette, France}
\author{T. Nishimichi}
\affiliation{Kavli Institute for the Physics and Mathematics of the Universe (WPI),
UTIAS, The University of Tokyo, Kashiwa, Chiba 277-8583, Japan}
\author{M. Rizzato}
\affiliation{Sorbonne Universités, CNRS, UMR 7095, Institut d'Astrophysique de Paris, 98 bis bd Arago, 75014 Paris, France}
\affiliation{Sorbonne Universités, Institut Lagrange de Paris (ILP), 98 bis bd Arago, 75014 Paris, France}

\begin{abstract}
We use a simple cosmological model with two parameters $(A_s, n_s)$ to illustrate the impact of using Halofit on error forecast based on Fisher information matrix for a $h^{-3} Gpc^3$ volume survey. We show that Halofit fails to reproduce well the derivatives of the power spectrum with respect to the cosmological parameters despite the good fit produced for its amplitude. We argue that the poor performance on the derivatives prediction is a general feature of this model and we exhibit the response function for the Halofit to show how it compares with the same quantity measured on simulations. The analytic structure of the Halofit response function points towards the origin of its weak performance at reproducing the derivatives of the non-linear power spectrum, which translates into unreliable Fisher information matrices.
\end{abstract}

\maketitle 

Considering a gas of cold dark matter (CDM) particles, we can study the problem of large scale structure formation on linear and mildly non-linear scales \cite{peebles1980large, bernardeau2002large}. A system of coupled, non-linear differential equations composed of continuity, Euler and Poisson equations will describe the evolution of density fluctuations and velocity of the CDM fluid under the action of a gravitational potential. As long as the system can be linearized, different Fourier modes evolve independently, and the signature of time evolution is factorized in the growth factor. Linearization, however, is based on the assumption of small density fluctuations and velocities. If the gravitational attraction prevails over the background space-time expansion, density fluctuations are no longer small as sign of structure formation, and the justification for the linearization of the dynamical equations disappears. The non-linear terms on the equations of motion produce couplings among different Fourier modes, and the power spectrum for a given wave mode will depend on all the other modes. 

\begin{figure}[!ht]
\centering
\includegraphics[width=8.5cm]{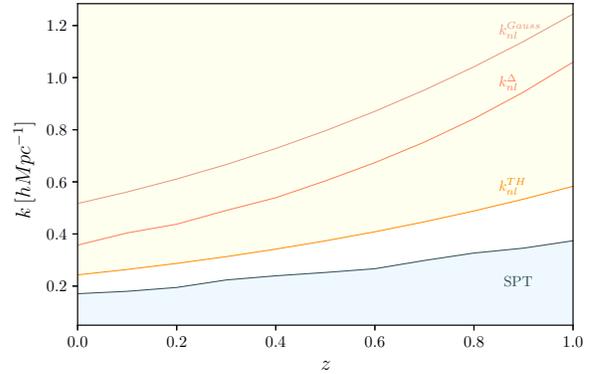}
\caption{Delimitation of the domain of non-linear regime as a function of redshift. The region where 1-loop Standard Perturbation Theory (SPT) compares with simulated data with errors less than $1\%$~\cite{jeong2006perturbation} is indicated by the lower shaded region (in green/blue), and the translinear region is the upper shaded areas (in yellow). Three alternative criteria can be used to define the non-linear region, based on Eqs.~\eqref{deltak} or \eqref{sigma_filter}. We use a power spectrum in a $\Lambda CDM$ cosmology with $n_s = 0.96$, $h=0.701$, $\Omegam = 0.279$, and $\sigma_8^2=0.815$.}
\label{scales}
\end{figure}

The transition from linear to non-linear regime is not sharply determined in the theory, and indicators of the reach of the non-linear regime may be given by the shape of the power spectrum or by the rms amplitude of mass fluctuations inside some spherical window. Considering the dimensionless quantity constructed in terms of the linear power spectrum:
\begin{equation}
\label{deltak}
\Delta^2(k) = \frac{k^3 P(k)}{2 \pi^2} \, ,
\end{equation}
we can define a scale of transition $k_{nl}^\Delta$ as the value of $k$ such that $\Delta^2(k_{nl}^\Delta) = 1$. Alternatively, the non-linear transition can be defined through the condition $\sigma^2_{1/k_{nl}} = 1$, where
\begin{equation}
\label{sigma_filter}
\sigma^2_R = \int \frac{\dd k k^2}{2 \pi^2} P(k) W(kR)^2
\end{equation}
and $W$ is a filter. If we take a Gaussian filter as $W$, we will have $k_{nl}^{Gauss}$. For a top-hat filter, we have $k_{nl}^{TH}$. The regions delimited by $k_{nl}^\Delta$, $k_{nl}^{Gauss}$, and $k_{nl}^{TH}$ are shown in Fig.~\ref{scales} as function of $z$ for a $\Lambda CDM$ cosmology.

All these scales, even if not exactly close to each other, give an idea of transition from an almost homogeneous and isotropic fluid subjected to small perturbations to a regime of halo formation, and halo models seem promising approaches to handle such non-linear domains \cite{cooray2002halo}. At an intermediate scale between the linear and the non-linear regimes, the mildly non-linear scales can be treated analytically by perturbation theory \cite{valageas2001dynamics}. Comparison between the predictions of the matter power spectrum from Standard Perturbation Theory \cite{bernardeau2002large} (SPT)  at 1-loop and simulated data were performed, and a scale indicating an agreement better that $1\%$ is proposed as $\Delta^2(k) \leq 0.4$ \cite{jeong2006perturbation, carlson2009critical}. This region of agreement is also shown in Fig. \ref{scales}.

The exploration of the mildly non-linear regime conducts indeed to a large gain in terms of cosmological information. Studies of the information content of the non-linear power spectrum indicate that the cumulative information grows with $k$ until the reach of the translinear regime (yellow shaded region on Fig. \ref{scales}), where a plateau starts \cite{rimes2005information, neyrinck2006information, neyrinck2007information, lee2008information, sato2009simulations, carron2015information}. The white region of Fig. \ref{scales} contains, therefore, rich information and mining it will be necessary for a proper analysis of the data from large galaxy surveys. 

\begin{figure}[!ht]
\centering
\includegraphics[width=9cm]{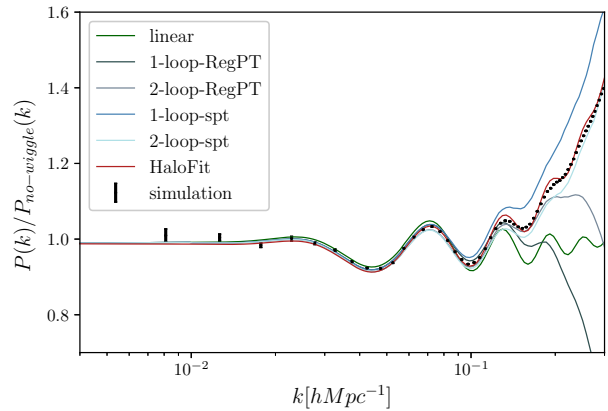}
\caption{Comparison between the prediction for the dark matter power spectrum using linear theory, 1-loop SPT, 2-loop SPT, 1-loop RegPT, 2-loop RegPT,  HaloFit and simulated data from~\cite{taruya2012direct} at $z=0.35$.}
\label{models}
\end{figure}

Perturbation theories are based on fundamental fluid dynamics and give reliable results for the power spectrum at mildly non-linear scales. Standard Perturbation Theory (SPT) and resumed versions such as RegPT \cite{taruya2012direct} have no free parameters and numerical codes are available for computations at 2-loop level. Faster predictions for the power spectrum amplitude can be obtained from halo model inspired fit formulas like Halofit
 \cite{cooray2002halo, smith2003stable, ma2000deriving, scoccimarro2001many, takahashi2012revising}. Halofit parametrizes the non-linear effects as a sum of a term related to the small scales interactions inside a halo -- the one-halo term -- and the interactions between different halos -- the two-halo term -- and the full model is fitted using numerical simulations. It is however a fit formula: equations with a form predicted from halo models have a large number of free parameters that are fixed using n-body simulations and despite its success, it has some weaknesses such as the underestimation of the power spectrum on the transition from two-halos to one-halo regimes \cite{valageas2011combining}. Performance tests of different perturbative schemes were performed in the literature \cite{carlson2009critical} but there is a large class of perturbation techniques relying on different assumptions and a full panorama of the field is hard to give. We show in Fig. \ref{models} a comparison between the predictions for the matter power spectrum produced by SPT at 1 and 2 loops, RegPT at 1 and 2 loops \cite{taruya2012direct}, and the HaloFit fit formula \cite{takahashi2012revising}. 

In what follows we will consider Halofit as presented in \cite{takahashi2012revising} to show that, despite the good agreement on the amplitude fitting, its performance isn't at the same level of accuracy on predicting the derivatives of the non-linear power spectrum with respect to the cosmological parameters. In order to measure the impact of the poor prediction for the derivatives of the non-linear power spectrum produced by Halofit, we will consider a two-parameter cosmological parameter on section \ref{two-parameters} and show how error forecast based on Fisher information matrix analysis could be affected. We will then argue on section \ref{response_func} that the behavior of the derivatives displayed on the specific two-parameter model considered is indeed intrinsic to the Halofit model and rooted on its failure at reproducing the response functions measured on the simulations.

\section{Two-parameter model}
\label{two-parameters}

\begin{figure}[!ht]
\centering
\includegraphics[width=9cm]{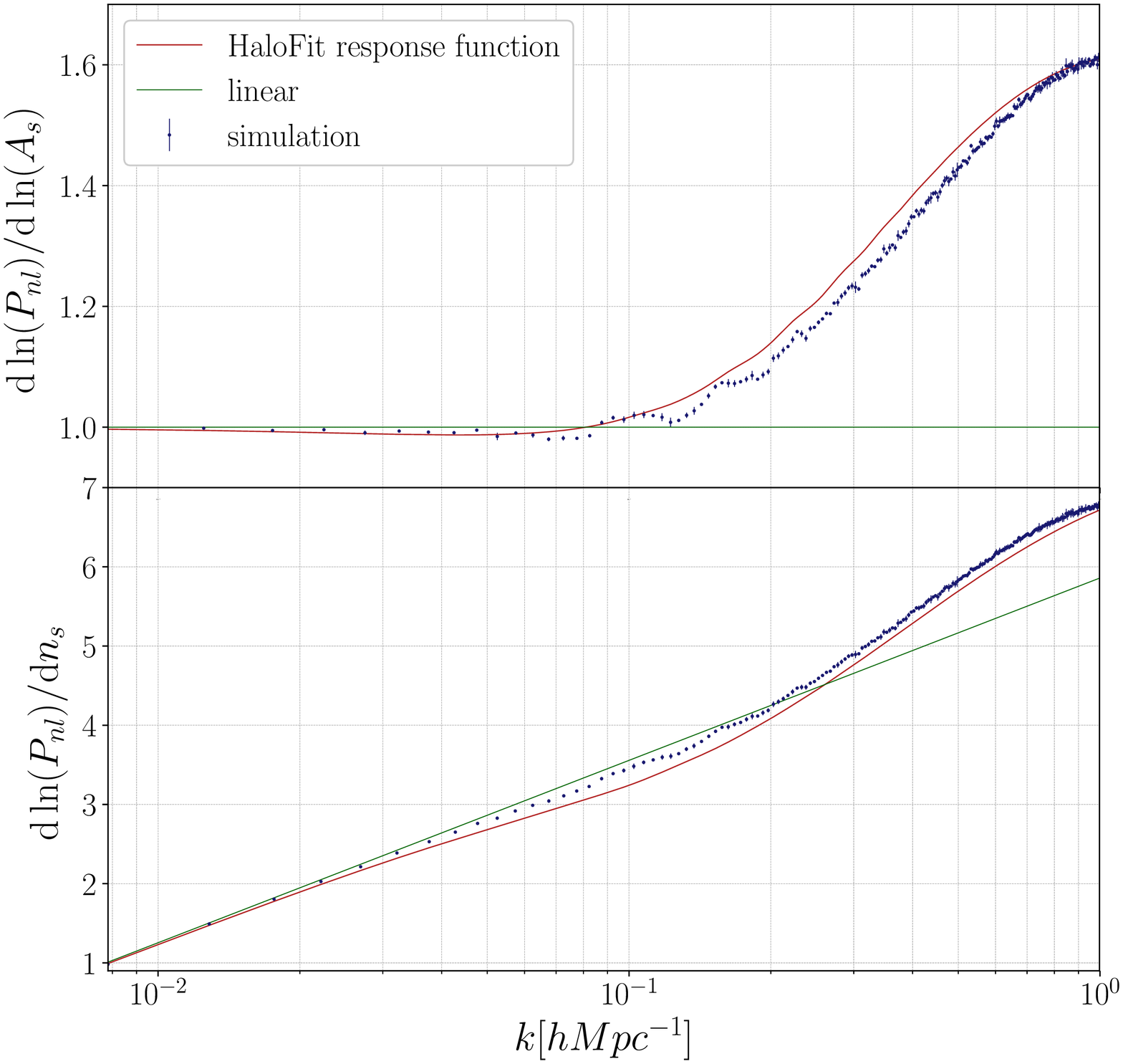}
\caption{The plotted curves correspond to the log derivative of the non-linear power spectrum wrt $A_s$ and $n_s$ given by Halofit at $z=0.35$. The red points correspond to the data measured from simulation. }
\label{ds_z035}
\end{figure}

In order to examine the performance of Halofit at predicting the derivatives of the non-linear power spectrum with respect to the cosmological parameters, and its impact on a Fisher matrix forecast based on Halofit, we consider a very simple cosmological model where only $A_s$ and $n_s$ are allowed to vary. We assume here $\Lambda CDM$ model with $\Omegam = 0.279$, $n_s = 0.96$, $h=0.701$,  and $\sigma_8^2 = 0.815$ and $k_0=0.002 Mpc^{-1}$.
We display first the results for $\frac{\dd \ln( \Pnl (k))}{\dd \ln(A_s)}$ and $\frac{\dd \ln( \Pnl (k))}{\dd n_s}$ at $z=0.35$ in Fig. \ref{ds_z035}. This shows the relative behavior of the measured data from simulation, the linear theory and the prediction from HaloFit model.

In order to obtain the points in Fig. \ref{ds_z035}, we have performed four pairs of $N$-body simulations employing $512^3$ particles in a $1\,h^{-3} Gpc^3$ periodic comoving cube with slightly varied cosmological parameters for both $A_s$ and $n_s$ to estimate the derivative based on a finite difference method ($16$ simulations in total). We vary $A_s$ ($n_s$) by $\pm 1\%$ ($\pm 0.01$) from their fiducial values for the two simulations in a pair and they share the initial random phases such that the cosmic variance error is largely cancelled when we take the numerical derivative. The initial conditions are generated using a code based on second-order Lagrangian Perturbation Theory \cite{2006MNRAS.373..369C,2009PASJ...61..321N}. Then the particle distribution is evolved by a Tree-PM code \texttt{Gadget2} \cite{Springel:2001,Springel:2005}.

We measure the matter power spectrum from the snapshots at $z=0.35$ using the standard method based on Fast Fourier Transform with $1024^3$ grid points and the Cloud-in-Cells interpolation scheme \citep{hockney81}. We then take the difference of the measured power spectra between a simulation pair to have an estimate of the derivative of $P^\mathrm{nl}(k)$ with respect to $A_s$ and $n_s$. Finally, the mean and the scatter of the estimates among the four pairs are recorded as our final estimate and uncertainty.

Note that the simulation settings described above are consistent with those in \cite{nishimichi2017moving}, from which we take the numerical values of the response function (see Figs.~\ref{kernel1}--\ref{kernel3} below).

The recent work \cite{smith2018precision} also presents the comparison between Halofit predictions and measurement from simulation of derivatives of the power spectrum with respect to eight cosmological parameters. The discrepancy presented in Fig. \ref{ds_z035} is not a particular feature of our choice of cosmological parameters.

\subsection{Fisher Matrix}

Fisher information matrix provides important boundaries on the variances associated to the measured parameters on an experiment 
and allows to bypass a full MCMC exploration of the space of parameters allowed by the observations. It is defined as \cite{tegmark1997measuring}:
\begin{equation}
\label{Fisher}
\mathbf{F}_{\alpha \beta} = \left \langle \frac{\partial^2 \ln \mathcal{L}}{\partial \alpha \partial \beta} \right \rangle
\end{equation}
where $\mathcal{L}$ is the likelihood, i.e., the probability of the data given a model $\Theta$ with parameters $\alpha, \beta$.

Given that our observable is the 3D matter power spectrum, if we assume that its covariance matrix does not vary that much with the parameters of the model, once we marginalize over the measured modes up to $k_{max}$ Eq.~\eqref{Fisher} specifies to \cite{tegmark1997measuring}:
\begin{equation}
\label{Ourfisher}
\mathbf{F}_{\alpha \beta} = \sum_{k_i, k_j \leq k_{max}} \frac{\partial P(k_i)}{\partial \alpha} \mathrm{Cov}_{ij}^{-1} \frac{\partial P(k_j)}{\partial \beta} \, .
\end{equation} 
The two key ingredients in Eq.~\eqref{Ourfisher} are the derivatives of the power spectrum (observable) and its covariance matrix (error budget). The determination of the covariance matrix requires in general the modeling of the four point correlation function of the dark matter field. Since we are only interested on the general behavior of the Fisher matrix, we will rely on  the estimation of the covariance matrix proposed in \cite{carron2015information}
\begin{equation}
\label{CovMat}
\text{Cov}_{ij} = \delta_{ij}\frac{2\left(P\left(k_i\right) + \frac{1}{\bar{n}}\right)^2}{N_{k_i}} + \sigma_{\text{min}}^2P\left(k_i\right)P\left(k_j\right)
\end{equation}
where $k_i$ is the representative value for the momenta in the $i^{th}$ bin, $\bar{n}$ is the shot noise term, $N_{k_i}$ is the number of independent $\mathbf{k}$ available for the estimation of the power spectrum in the bin $i$ and $\sigma_{min}^2$ can be estimated through general arguments related to hierarchical models. 

It follows from \eqref{Ourfisher} and \eqref{CovMat} that the Fisher matrix for two parameter $\alpha, \beta$ can be analytically expressed as
\begin{equation}
\mathbf{F}_{\alpha \beta}  = \mathbf{F}_{\alpha \beta}^{G} - \sigma_{min}^2 \frac{\mathbf{F}_{\alpha \ln A_s}^{G} \,  \mathbf{F}_{\ln A_s \beta}^{G} }{1 + \sigma^2_{min} \, \mathbf{F}_{\ln A_s \ln A_s}^{G}}
\end{equation}
where $\mathbf{F}_{\alpha \beta}^{G}$ is the classical gaussian Fisher information matrix derived in \cite{tegmark1997measuring}
\begin{equation}
\mathbf{F}_{\alpha \beta}^{G} =  \frac{V}{2 \pi^2} \int_{k_{min}}^{k_{max}} \dd k k^2 \left[ \frac{P(k)}{P(k) + 1/\bar{n}} \right]^2 \frac{\partial \ln P}{\partial \alpha} \frac{\partial \ln P}{\partial \beta} \, .
\label{fisher_gauss}
\end{equation}
The values for a $1 h^{-3} Gpc^3$ volume, we take $\bar{n} = 3 \, 10^{-4} h^3 Mpc^{-3}$ and $\sigma^2_{min} = 1.5 \, 10^{-4}$ as in \cite{carron2015information}. This form of the covariance matrix is convenient for its simplicity and also for incorporating non-gaussian corrections to the analysis. The fundamental fact for us in this paper, however, is that the covariance matrix provides a metric for the comparison of the derivatives of the power spectrum computed through distinct procedures. Our results are therefore not strongly dependent on the form and parametrization of the covariance matrix, but rely mostly on the fact that a positive definite matrix is uniquely defined for all possible sets of models for which the derivatives of $P(k)$ are compared.

\begin{figure}[!ht]
\centering
\includegraphics[width=8.5cm]{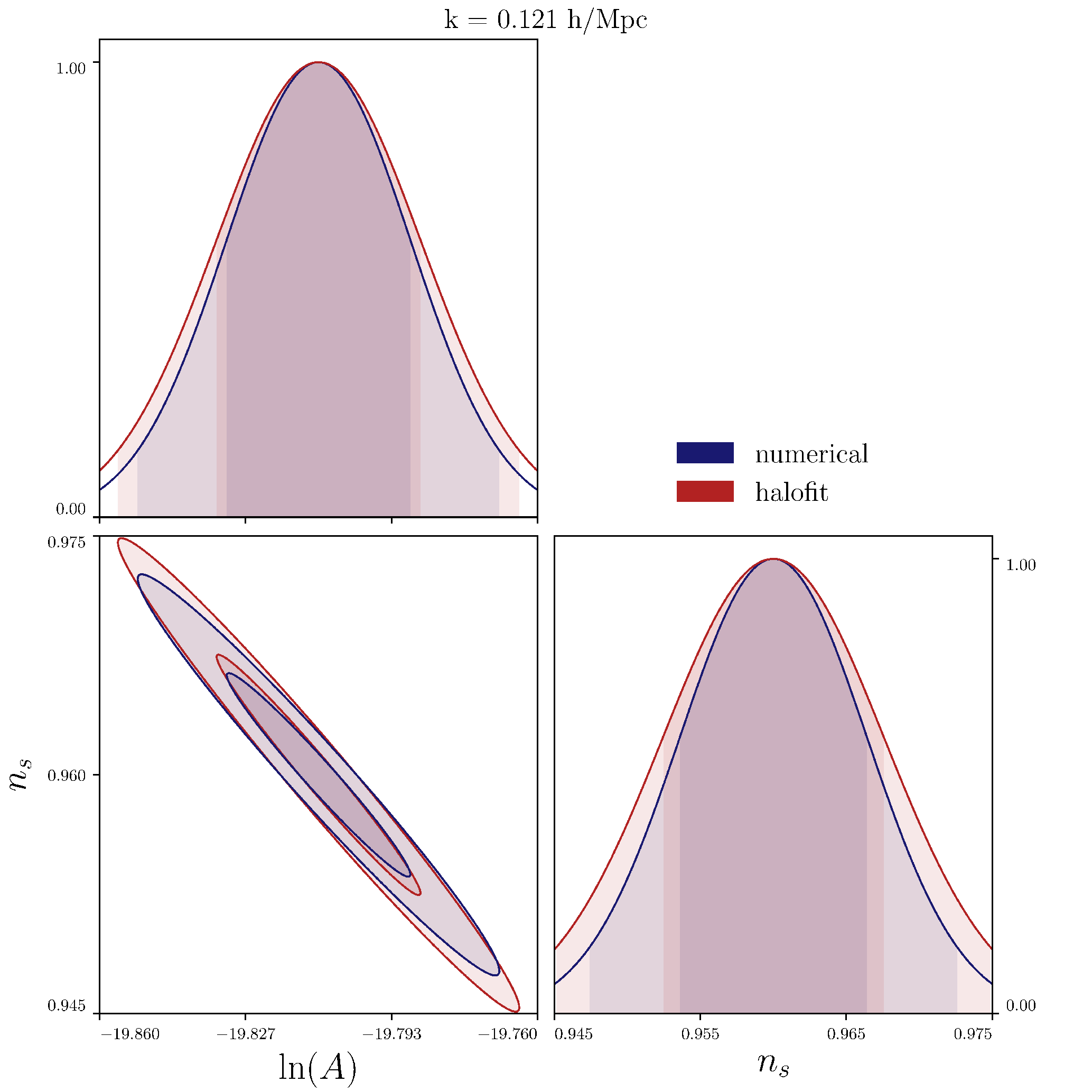}
\caption{Ellipses for numerical and Halofit Fisher matrices at $k=0.121 h/Mpc$ at $z=0.35$.}
\label{ellipses}
\end{figure}

Following this philosophy, we will build two families of Fisher matrices indexed by $k$ ($=k_{max}$ in Eq.~\eqref{fisher_gauss} and $k_{min} = 0.008 h Mpc^{-1}$): $\mathbf{F}_{ij}^N$ will be constructed using the measured values of $\frac{\partial \ln P}{\partial n_s}$ and $\frac{\partial \ln P}{\partial \ln A_s}$ on the simulation, and $\mathbf{F}_{ij}^H$ will use the prediction of Halofit for the same derivatives. As an illustration, the posteriors and the joint posterior for the two models at $z=0.35$ and $k=0.121 h/Mpc$ are compared in Fig.~\ref{ellipses}. The ellipses are generated with Cosmicfish\footnote{\texttt{cosmicfish.github.io}} \cite{raveri2016cosmicfish}.

The comparison of the ellipses produced by $\mathbf{F}^N$ and $\mathbf{F}^H$ for all $k$s can be performed by looking at two quantities: their relative area and relative inclination. The ratio of the area of the ellipses is given by the ratio of their determinants, or in other terms, the ratio of the Figures of Merit (FoM). For the relative inclination we take the ratio of the angles determined by the eigenvectors of each matrix. Namely, defining $\phi_{\, \cdot \, } = \arctan(v_2^{\, \cdot \,}/v_1^{\, \cdot \,})$, where $\{ v_1^{\, \cdot \,}, v_2^{\, \cdot \,} \}$ are the eigenvectors of $\mathbf{F}_{ij}^{\, \cdot \,}$, we can have a measure of the relative inclination as $\phi_N/\phi_H$. 
\begin{figure}[!ht]
\centering
\includegraphics[width=9cm]{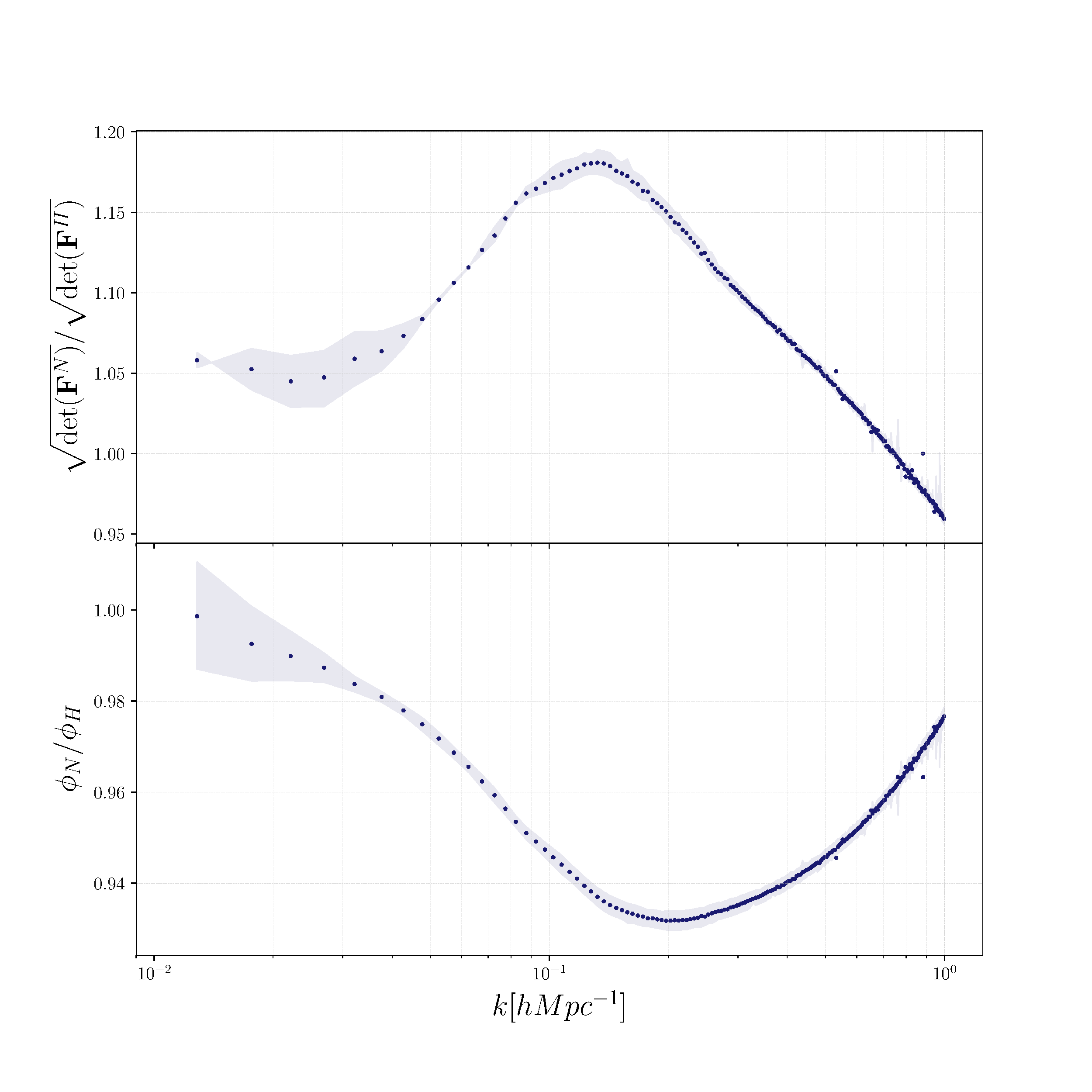}
\caption{Ratio of the figures of merit and relative inclination for Fisher information matrices constructed using numerical data and Halofit at $z=0.35$. The parameters used for the covariance matrix are those given in \cite{carron2015information}. The shaded region correspond to reconstructions based on the numerical value of the derivatives plus or minus their corresponding error.}
\label{FoM}
\end{figure}

We show in Fig.~\ref{FoM} the results for the relative area and relative inclination for error ellipses as function of $k$. Comparing the results of Fig.~\ref{FoM} with the shaded regions in Fig.~\ref{scales} we realize that Halofit performs poorly in the regime where SPT is precise at percent level (i.e., $k \lesssim 0.2\,h^{-1}Mpc$; see Fig.~\ref{models}), indicating that a mixed model between perturbation theory and Halofit could be better suited for Fisher error matrix purposes. 

We stress here that the departure of the Halofit predictions from the simulated ones is not due to the numerical computation of derivatives, and indeed these derivatives were computed through the response function as presented in Sec.~\ref{response_func}. The comparison between numerical derivatives and the ones produced through the response function is performed at a high level of accuracy. 

We claim therefore that the discrepancies present in Fig.~\ref{FoM} -- that could be antecipated from Fig.~\ref{ds_z035} --  are not related to the cosmological parameters chosen in this analysis nor to the particular calculation procedure, but rather a general result due to the poor performance of Halofit on the prediction of the derivative of the power spectrum with respect to the cosmological parameters. We will show in the next section that the root of this behavior is related to the response function produced by Halofit.

\section{Response functions}
\label{response_func}

The predictions from different perturbation theories and fit formulas in the mildly non-linear regime should be systematically compared based on internal consistency and quality of the approximation to simulated data. A second level of inquiry lies in the analysis of the response functions. A response function, as defined in~\cite{nishimichi2016response}, is the functional derivative of a prescribed non-linear power spectrum with respect to the linear power spectrum that generated it. The response functions can be measured on simulations and compared with analytic results \cite{nishimichi2016response, nishimichi2017moving}. Explicitly, at a fixed redshift, it expresses as
\begin{equation}
\label{response_def}
R(k, q) = \frac{\delta \Pnl (k)}{\delta \Plin (q)} \, .
\end{equation}

The clear interest on response functions for the non-linear power spectrum on the context of Fisher matrix error forecast comes from an application of the chain rule: the derivative of the non-linear power spectrum with respect to a cosmological parameter can be expressed as the convolution of the response function and the derivative of the linear power-spectrum with respect to the cosmological parameter of interest ($\theta$ generically):

\begin{equation}
\label{chain_rule}
\frac{\dd \Pnl (k)}{\dd \theta} = \int \dd q \frac{\delta \Pnl (k)}{\delta \Plin (q)} \frac{\dd \Plin (q)}{\dd \theta} \, ,
\end{equation}
where the numerical derivatives of $\Plin (q)$ can be computed via Einstein-Boltzmann codes such as CAMB\footnote{\texttt{camb.info}} or CLASS\footnote{\texttt{class-code.net}}.

We should look at response functions both as a mean to compute the non-linear power spectrum, its derivatives, and also as possible mean of diagnosing issues on predictions for the amplitude of the non-linear corrections  and, in particular, test the robustness of HaloFit at this level.

In order to illustrate a general feature of the response functions considered in this work and fix notation for decompositions on which our future arguments will be based, let us consider the response function for standard perturbation theory as prototype. In SPT the non linear power spectrum can be written, at 1-loop, as \cite{2002PhR...367....1B, 2008PhRvD..78j3521B}:
\begin{align}
 \Pspt(k) &= (1 + 2 \Gamma^{(1)}_{1-loop}(k)) \Plin(k) \nonumber\\
 &+ 2 \int \frac{\dd^3 \vq}{(2 \pi)^3} [F_{sym}^{(2)}(\vq, \vk - \vq)]^2 \times \\
 &\hspace{1cm}\times \Plin(|\vk - \vq|) \Plin(q)\nonumber \, .
\end{align}
Not focusing at the nature of each term, but on the general structure of the equation, we can easily recognize a term proportional to $\Plin(k)$ and a second term with more complicated structure in which different modes are coupled. In the same way, the power spectrum from Halofit parametrization can be factorized as the sum of a one-halo and a two-halos terms, the latter being proportional to $\Plin$ as shown in Eq.~\eqref{deltaNL}. By taking the functional derivative of a non-linear power spectrum with this structure to construct $R(k, q)$ as in Eq. \eqref{response_def}, we obtain generically for these models:
\begin{equation}
\label{response_decomposed}
R(k, q) = \rdelta (k) \,  \Dirac(k-q) + \rsmooth (k, q) \, .
\end{equation}
For convenience we will define the kernel $K(k, q)$ as  constructed from the smooth contribution to the response function
\begin{equation}
\label{Kdef}
K(k, q) = q \, R^{\mathrm{smooth}}(k, q) \, .
\end{equation}
We observe that \cite{nishimichi2016response} defines $K(k, q) = q \, R(k, q)$, but we want to focus on the smooth part only. 

\subsection{The measured kernels}
The response function can be measured from simulations \cite{nishimichi2016response} and computed from theoretical models for the non-linear growth of structures. Therefore we can test the agreement between theory and N-body simulations not only at the level of the power spectrum but also at the level of the response functions. The comparison between response functions measured from simulations and those predicted from SPT and RegPT are studied in \cite{nishimichi2016response, nishimichi2017moving}. The response functions can also be computed for the Halofit model, and the derivation is presented in App.~\ref{response_computation}.

\begin{figure}[!ht]
\centering
\includegraphics[width=8.5cm]{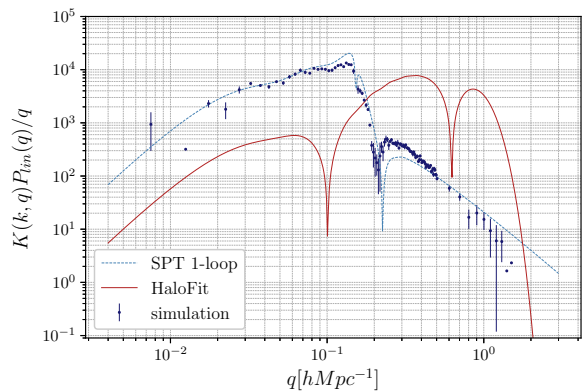}
\caption{SPT-1loop, Halofit and measure points at $z=0.35$. The points correspond to the dataset for the kernel measured at $k=0.1525 h Mpc^{-1}$.}
\label{kernel1}
\end{figure}

\begin{figure}[!ht]
\centering
\includegraphics[width=8.5cm]{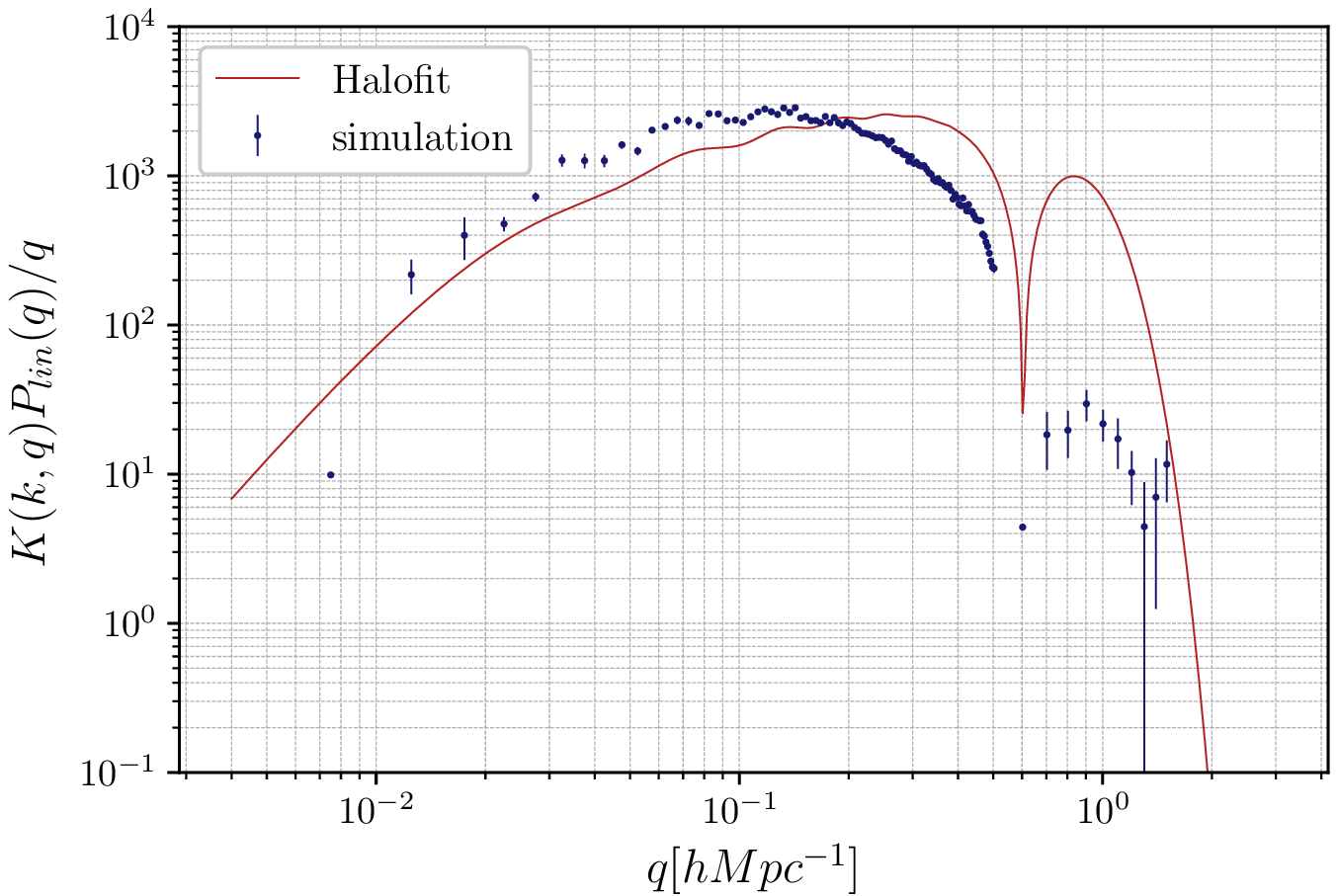}
\caption{Halofit and measure points at $z=0.35$. The points correspond to the dataset for the kernel measured at $k=0.4525 h Mpc^{-1}$.}
\label{kernel2}
\end{figure}

\begin{figure}[!ht]
\centering
\includegraphics[width=8.5cm]{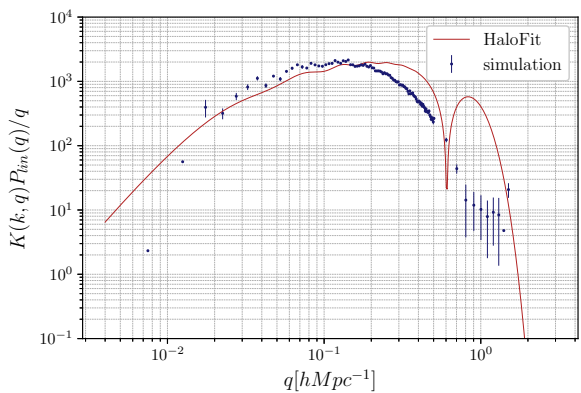}
\caption{SPT-1loop, Halofit and measure points at $z=0.35$. The points correspond to the dataset for the kernel measured at $k=0.6025 h Mpc^{-1}$.}
\label{kernel3}
\end{figure}

In Fig.~\ref{kernel1}, \ref{kernel2} and \ref{kernel3} the kernels $K(k, q)$ are plotted as function of $q$ at $z=0.35$ for $k= 0.1525 \, h Mpc^{-1}$, $k=0.4525 \, h Mpc^{-1}$ and  $k=0.6025 \, h Mpc^{-1}$, respectively. Even if SPT is no longer supposed to predict well the response function at $k= 0.1525 \, h Mpc^{-1}$ for the redshift considered, we trace its response at one loop in Fig.~\ref{kernel1} just to display its global behavior. We can see that even beyond its regime of validity, SPT reproduces the general features of the data points.

The disagreement between measured kernels and Halofit predictions is more dramatic for small $k$s. In order to investigate the reasons for this departure from the measured points we can combine Eqs.~\eqref{chain_rule} and \eqref{response_decomposed} to compute the following quantity:
\begin{eqnarray}
\label{d_difference}
& & \PH(k) \frac{\dd \ln \PH}{\dd \ln A_s}(k) - \Pspt(k) \frac{\dd \ln \Pspt}{\dd \ln A_s}(k) \nonumber\\ & & = \left[\rdelta_{\mathrm{Hf}}(k) - 
\rdelta_{\mathrm{SPT}}(k)\right] \Plin(k) \nonumber\\ & &
+ \int \dd q \, \left[ \rsmooth_{\mathrm{Hf}}(k, q) - \rsmooth_{\mathrm{SPT}}(k, q) \right] \Plin(q) \, .
\end{eqnarray}
The derivative with respect to the amplitude $A_s$ is taken for simplicity. The indexes $\mathrm{Hf}$ stand to Halofit. The first and second terms on the rhs of Eq.~\eqref{d_difference} are the differences of the $\rdelta$ and $\rsmooth$ using the nomenclature given in Eq. \eqref{response_decomposed}.
\begin{figure}[!ht]
\centering
\includegraphics[width=8.5cm]{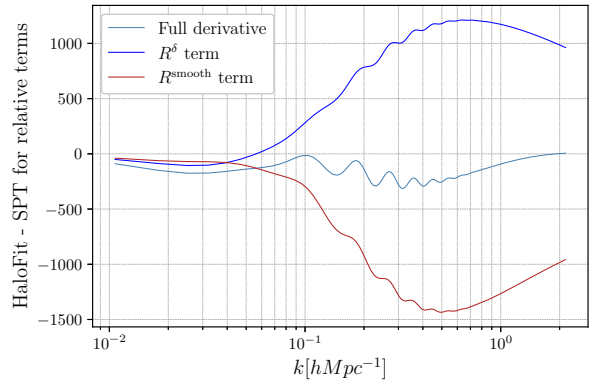}
\caption{Behavior of the lhs and of each of the terms on the rhs of Eq. \eqref{d_difference} at $z=0.35$.}
\label{relative_terms}
\end{figure}

\begin{figure}[!ht]
\centering
\includegraphics[width=8.5cm]{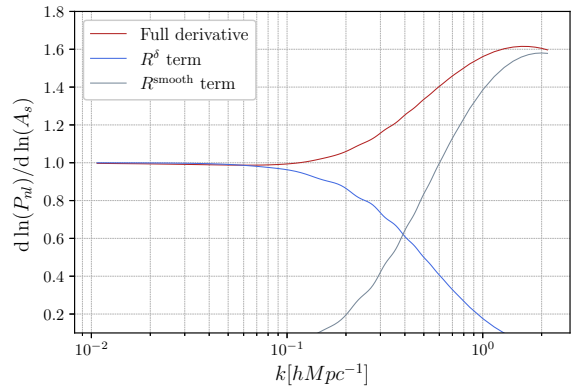}
\caption{Contributions of the $\rdelta$ and $\rsmooth$ to $\frac{\dd \ln \PH}{\dd \ln A_s}(k)$ at $z=0.35$.}
\label{halofit_contributions}
\end{figure}

We observe that the derivatives of the non-linear power spectrum computed with Halofit and 1-loop SPT are not coincident (as expected) but their difference remains limited to a region around zero, oscillating but with no tendency to grow or decrease in this range of $k$s,  as shown in Fig. \ref{relative_terms}. The same is not true for the $\rdelta$ and $\rsmooth$ terms: they grow in opposite directions but almost compensate each other. The Halofit $\rdelta$ term is normalized to be in accord with halo profiles and has a slow decrease if compared with SPT \cite{crocce2008nonlinear}. This implies that $\rdelta$ dominates the response function at small $k$s, as we can clearly see from Fig. \ref{halofit_contributions}. The smooth component of the response function is then constrained to contribute much less that its SPT counterpart at this regime. Since the SPT $\rsmooth$ term reproduces the global behavior of the measured data points, we should not be surprised by desagreament between Halofit predictions and measured datapoints displayed in Fig. \ref{kernel1}.

The relative contributions of $\rdelta$ and $\rsmooth$ terms for $\frac{\dd \ln \PH}{\dd \ln A_s}(k)$, shown in Fig. \ref{halofit_contributions}, indicates that the $\rsmooth$ component start dominating the response function on non-linear scales, what explains the proximity of the data points to the Halofit kernels presented in Figs. \ref{kernel2} and \ref{kernel3}, even though the high $q$ behavior is not reproduced.

\subsection{Leakage of the one-halo term}

The response function measures how much the non-linear power spectrum at a given scale $k$ is affected by a change in the linear power spectrum at a given scale $q$. As discussed, the response function can be decomposed as $R(k, q) = \rdelta (k) \,  \Dirac(k-q) + \rsmooth (k, q)$ and $\rsmooth$ can be further decomposed into a one-halo and a two-halos contribution, as in Eqs. \eqref{deltaNL} and \eqref{2halos}. In this framework, the one-halo term encodes the effect of the physics concentrated on scales around and smaller than the size of a halo while the two-halo term should describe interaction between halos i.e. wide range interactions. Let $\mathcal{R}_{1H} (k)$ be defined as
\begin{equation}
\mathcal{R}_{1H} (k) := \frac{\int \dd q \, K_{1H}(k, q) \Plin(q)}{\int \dd q \, K(k, q) \Plin(q)}
\end{equation}
where $K_{1H}(k, q)$ is the corresponding contribution in Eq. \eqref{Kdef} due to the 1-halo term on $\rsmooth$.

\begin{figure}[!ht]
\centering
\includegraphics[width=9cm]{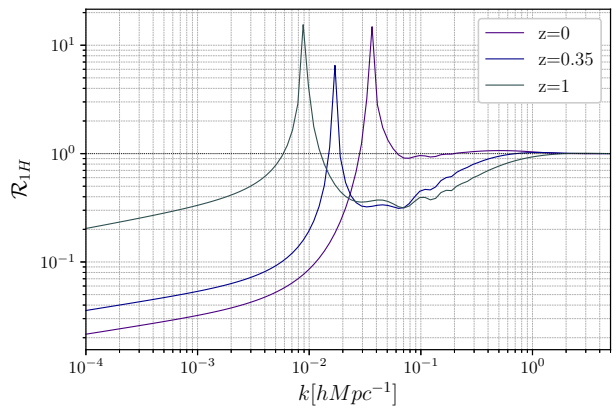}
\caption{Contribution from the 1H term to the total area delimited by the kernel $K$.}
\label{area}
\end{figure}

Given the interpretation of the 1-halo and the 2-halo terms, we should expect $\mathcal{R}_{1H} (k) \approx 1$ for high $k$s, what is indeed the case as shown in Fig. \ref{area}. For small $k$s we should expect a rapid decrease of $\mathcal{R}_{1H} (k)$, what is not verified.

Even if the computation of derivatives of the non-linear power spectrum through the response function will be dominated by the $\rdelta$ term at small scales, $\mathcal{R}_{1H} (k)$ ranging around the percent level at small $k$s indicate a leakage of the 1-halo term to extremely large scales, what is unphysical and due to the choice of parametrization functions and fitting process.

\section{Conclusion}

Halo models parameters encapsulate the general features of the power spectrum but they are not capable to grasp the full gravitational physics acting on modes coupling and structure formation. Fitting the power spectrum  based on halo model parametrizations such as Halofit constraint the model to reproduce the amplitude under certain regimes, but will not necessarily reproduce also its derivatives with respect to all cosmological parameters. We must therefore be cautious on performing error forecasts based on Fisher information matrices and using these models.

The discrepancy between the derivatives measured on simulation and the predicted by Halofit is rooted on the poor behavior of the Halofit response function when compared with the numerical points, most importantly at low $k$. We can directly point at the leakage of the one-halo term into large scales as an unphysical feature.

We took a two-parameter cosmological model Fisher information matrix to illustrate a particular feature of more general phenomenon: since the response function connects derivatives of the linear and non-linear power spectrum with respect to cosmological parameters through a simple chain rule, the estimation of all derivatives and construction of Fisher information matrices for a larger set of parameters are faded to suffer from the same pathology. Indeed we can see in Fig 11 from \cite{smith2018precision} that Halofit performs poorly for the global set of parameters used in their analysis.

Two lessons come out of this exercise: first, error forecasts based on Fisher information matrices that use fitted non-linear power spectrum such as Halofit may hide subtle effects and inaccuracies if the response function is not well reproduced by the non-linear model. Secondly, tests of fit formulas and emulators should also aim at comparing their response functions with the response functions measured on the simulations.

\section{Acknowledgments} 

This work was partially supported by Grant No. ANR-12-BS05-
0002 and Labex ILP (reference ANR-10-LABX-63) part of
the Idex SUPER of the program Investissements d'avenir
Grant No. ANR-11-IDEX-0004-02. 
TN is supported in part
by World Premier International Research Center Initiative (WPI Initiative), MEXT, Japan, and JSPS KAKENHI Grant Number 17K14273, and JST CREST Grant Number JPMJCR1414. Numerical simulations were carried out on Cray XC50 at the Center for Computational Astrophysics, National Astronomical Observatory of Japan.
P.R. and M.R. acknowledge financial support from the Centre National d'Etudes Spatiales (CNES) fellowship program.

\bibliographystyle{h-physrev}
\bibliography{references}

\appendix
\section{The HaloFit response function}
\label{response_computation}

The response function for SPT has already been explored on the literature \cite{nishimichi2016response} and the general case of RegPT is also under study \cite{nishimichi2017moving}. The calculation of the response function for the HaloFit model has not been computed yet in the literature, and therefore we present the basis of the calculation here.

The functional form of the revised halofit model is taken from \cite{takahashi2012revising}, where the formulas provided by \cite{smith2003stable} are re-analysed. Defining
\begin{equation}
\Delta_L^2(k) = \frac{k^3 \Plin(k)}{2 \pi^2}
\end{equation}
we can compute the variance of the density field using a gaussian filtering:
\begin{equation}
\label{sigmaR}
\sigma^2(R) = \int \dd \ln k \,  \Delta^2_L(k)  \, \e^{-k^2 R^2} \, .
\end{equation}
A non-linear scale $\rnl$ is defined for a given $\Plin(k)$ by the implicit condition:
\begin{equation}
\label{sigmaRnl}
\sigma^2(\rnl) = 1 \, .
\end{equation}
Let $y := k \rnl$, $f(y) = y/4 + y^2/8$. The non-linear power spectrum is given by:
\begin{equation}
\label{deltaNL}
\Delta^2_{NL}(k) = \Delta_{1H}^2(k) + \Delta_{2H}^2(k)
\end{equation}
where $\Delta^2_{NL}(k) = k^3 \Pnl(k)/(2 \pi^2)$, and

\begin{eqnarray}
\label{one-halo}
\Delta^2_{1H}(k) & = &  \frac{a_n \, y^{3 f_1(\Omegam)}}{A \, B} \, ,
\end{eqnarray}
with $A = 1+\mu_n y^{-1} + \nu_n y^{-2}$ and $B=1+b_n \, y^{3 f_2(\Omegam)} + [c_n \, f_3(\Omegam) y]^{3 - \gamma_n}$ is the one-halo term. The two-halos term is given by
\begin{equation}
\label{two-halo}
\Delta^2_{2H}(k) = \Delta^2_L(k) \frac{(1+\Delta_L^2(k))^{\beta_n}}{1 + \alpha_n \Delta_L^2(k)} \e^{-f(y)} \, .
\end{equation}

The functions $f_1, f_2, f_2$ depend on $\Omegam$ only. The parameters $a_n, b_n, c_n, \alpha_n, \beta_n, \gamma_n, \mu_n, \nu_n$ (see \cite{takahashi2012revising}) are functions of the effective spectral index $\neff$ and the curvature $C$, defined as:
\begin{equation}
\label{neff}
\neff + 3 = - \frac{\dd \ln \sigma^2(R)}{\dd \ln R} \bigg\rvert_{\sigma=1} \, ,
\end{equation}

\begin{equation}
C = - \frac{\dd^2 \ln \sigma^2(R)}{\dd \ln R^2} \bigg\rvert_{\sigma=1} \, .
\end{equation}

To obtain the response function $R(k, q)$ for the HaloFit we have to solve the exercise defined on Eq. \eqref{response_def} for the non-linear power spectrum predicted by HaloFit on Eqs. \eqref{deltaNL}, \eqref{one-halo}, \eqref{two-halo}. Since all the parameters $a_n, b_n, c_n, \alpha_n, \beta_n, \gamma_n, \mu_n, \nu_n$ are functions of $\neff$ and $C$, and these two quantities are implicitly given in terms of $\Plin$ through Eqs. \eqref{sigmaR}, \eqref{sigmaRnl}, we have to provide expressions for $\frac{\delta \neff}{\delta \Plin}$ and $\frac{\delta C}{\delta \Plin}$. 

\begin{equation}
\frac{\delta \neff}{\delta \Plin} = \frac{(\neff + 3)}{\rnl} \frac{\delta \rnl}{\delta \Plin} - \rnl \frac{\delta}{\delta \Plin} \frac{\dd \sigma^2}{\dd \rnl}
\end{equation}

\begin{eqnarray}
\frac{\delta C}{\delta \Plin} & = & - \left[ \frac{(\neff + 3) - 2C}{\rnl} \right]   \frac{\delta \rnl}{\delta \Plin} \nonumber\\ & & - (2 \neff + 7) \rnl \frac{\delta}{\delta \Plin} \frac{\dd \sigma^2}{\dd \rnl} \nonumber\\ & & - \rnl^2 \frac{\delta}{\delta \Plin} \frac{\dd^2 \sigma^2}{\dd \rnl^2}
\end{eqnarray}

We must then compute $\frac{\delta \rnl}{\delta \Plin}$. For this end, we observe that Eq. \eqref{sigmaRnl} implies:
\begin{equation}
0 = \frac{\delta \sigma^2}{\delta \Plin(q)}  =   \frac{q^2}{2 \pi^2}  \e^{-q^2 \rnl^2}  + \frac{\delta \rnl}{\delta \Plin(q)}  \frac{\dd \sigma^2}{\dd \rnl} \, 
\end{equation}
and therefore
\begin{equation}
\frac{\delta \rnl}{\delta \Plin(q)} = \frac{q^2 \, \rnl \,  \e^{-q^2 \rnl^2}}{2 \pi^2 (\neff + 3)}
\end{equation}
where we used Eq. \eqref{neff} to express the derivative $\frac{\dd \sigma^2(\rnl)}{\dd \rnl}$ in terms of $\neff$.

We also have:
\begin{equation}
\frac{\delta}{\delta \Plin(q)} \frac{\dd \sigma^2}{\dd \rnl} = - \frac{q^4 \rnl}{\pi^2} \e^{-q^2 \rnl^2} +  \frac{\delta \rnl}{\delta \Plin(q)} \frac{\dd^2 \sigma^2(\rnl)}{\dd \rnl^2} \, ,
\end{equation}

\begin{eqnarray}
\frac{\delta}{\delta \Plin(q)} \frac{\dd^2 \sigma^2}{\dd \rnl^2} & = & \frac{q^4 (2 q^2 \rnl^2 -1)}{\pi^2} \e^{-q^2 \rnl^2} \nonumber\\ & & +  \frac{\delta \rnl}{\delta \Plin(q)} \frac{\dd^3 \sigma^2(\rnl)}{\dd \rnl^3} \, .
\end{eqnarray}
The derivatives $\frac{\dd^2 \sigma^2(\rnl)}{\dd \rnl^2}$, and $\frac{\dd^3 \sigma^2(\rnl)}{\dd \rnl^3}$ can be computed from the definition.

If we take the 2-halos term, the functional derivative gives:
\begin{eqnarray}
\label{2halos}
& & \frac{\delta \Delta^2_{2H}(k)}{\delta \Plin(q)}  =   \frac{k^3}{2 \pi^2} \frac{(1+\Delta_L^2(k))^{\beta_n}}{1 + \alpha_n \Delta_L^2(k)} \e^{-f(y)} \Bigg[ 1  + \frac{\beta \Delta_L^2(k)}{1 + \Delta_L^2(k)}  \nonumber\\ & & - \frac{\alpha \Delta_L^2(k)}{1 + \alpha \Delta_L^2(k)} \Bigg] \Dirac(k-q)  + \Delta^2_{L}(k) \nonumber\\ & & \times \frac{(1+\Delta_L^2(k))^{\beta_n}}{1 + \alpha_n \Delta_L^2(k)} \e^{-f(y)} \Bigg[  -\frac{k(1+k \rnl)}{4}  \frac{\delta \rnl}{\delta \Plin} \nonumber\\ & & +
\ln(1 + \Delta_L^2(k)) \left( \frac{\partial \beta}{\partial \neff} \frac{\delta \neff}{\delta \Plin} + \frac{\partial \beta}{\partial C} \frac{\delta C}{\delta \Plin} \right) \nonumber\\ & & - \frac{\Delta_L^2(k)}{1+\alpha \Delta_L^2(k)} \left( \frac{\partial \alpha}{\partial \neff} \frac{\delta \neff}{\delta \Plin} + \frac{\partial \alpha}{\partial C} \frac{\delta C}{\delta \Plin} \right) \Bigg] \, .
\end{eqnarray}
We see that this term has two contributions: a distributional component corresponding to a propagator term, and a smooth component corresponding to mode-coupling contributions.

The one-halo term does not involve $\Plin$ explicitly and therefore its functional derivative has no distributional component. Using the chain rule $\frac{\delta \Delta^2_{1H}(k)}{\delta \Plin(q)} $ can also be expressed in terms of $\frac{\delta \rnl}{\delta \Plin}$, $\frac{\delta \neff}{\delta \Plin}$, $\frac{\delta C}{\delta \Plin}$ and the partial derivatives of the parameters $a_n, b_n, c_n, \alpha_n, \beta_n, \gamma_n, \mu_n, \nu_n$ with respect to $\neff$ and $C$.


\end{document}